\documentclass[aps,prl,twocolumn,superscriptaddress,groupedaddress]{revtex4}  
\usepackage{graphicx}
\usepackage{amsmath}
\usepackage{float}
\usepackage{color}

\newcommand {\bbk}{{\bf k}}
\newcommand {\bbq}{{\bf q}}

\newcommand {\afo}{{\alpha^{2} F(\omega)}}

\begin{document}

\title{Electron-phonon interaction and pairing mechanism in superconducting Ca-intercalated bilayer graphene}
  
\author{E. R. Margine}
\email{rmargine@binghamton.edu}
\affiliation{Department of Physics, Applied Physics and Astronomy, 
Binghamton University - SUNY, Binghamton, New York 13902, USA}

\author{Henry Lambert}
\affiliation{Department of Materials, University of Oxford, Parks Road,
Oxford OX1 3PH, United Kingdom}

\author{Feliciano Giustino} 
\affiliation{Department of Materials, University of Oxford, Parks Road, 
Oxford OX1 3PH, United Kingdom}

\begin{abstract}
Using the {\it ab initio} anisotropic Eliashberg theory including Coulomb interactions, we investigate the electron-phonon interaction and the pairing mechanism in the recently-reported superconducting Ca-intercalated bilayer graphene. We find that C$_6$CaC$_6$ can support phonon-mediated superconductivity with a critical temperature $T_{\rm c}=6.8$--8.1~K, in good agreement with experimental data. Our calculations indicate that the low-energy Ca$_{xy}$ vibrations are critical to the pairing, and that it should be possible to resolve two distinct superconducting gaps on the electron and hole Fermi surface pockets.
\end{abstract}

\maketitle

Recent progress in the fabrication of metal-intercalated bilayer graphene \cite{Kanetani_PNAS12,Kleeman_PRB13,Kleeman_JPSJ14}, the thinnest limit of graphite intercalation compounds (GICs), opened promising new avenues for studying exotic quantum phenomena in two dimensions. Recently two independent studies reported the observation of superconductivity in Ca-intercalated bilayer graphene at 4~K~\cite{Hasegawa} and in Ca-intercalated graphene laminates around 6.4~K \cite{Chapman}, while a third study presented evidence for superconductivity in Li-decorated monolayer graphene around 5.9~K \cite{Ludbrook_PNAS15}. The electron-phonon interaction is expected to play a central role in these observations, hence it is important to develop a detailed understanding of electron-phonon physics in these newly-discovered superconductors.

The strength of the electron-phonon coupling (EPC) in graphite and graphene has been widely investigated using angle-resolved photoelectron spectroscopy (ARPES), however the interpretation of the results is not always straightforward~\cite{Bianchi_PRB10,Siegel_NJP12,Haberer_PRB13,Gruneis_PRB09,McChesney_PRL10,Fedorov_NatCom14,Yang_NatCom14}. For example the anisotropy of the EPC in alkali-metal decorated graphene generated significant debate in relation to the role of van~Hove singularities~\cite{Giustino_PRB08,Gruneis_PRB09,McChesney_PRL10,Fedorov_NatCom14}. Furthermore the relative importance of the $\pi^*$ bands and of the interlayer (IL) band in the pairing mechanism of bulk CaC$_6$ \cite{weller} has been the subject of an intense debate. Indeed several studies suggested that superconductivity arises from the EPC of both bands~\cite{Mazin_PRL05,Calandra_PRL05,Kim_PRL06,Boeri_PRB07,Sanna_PRB07,Yang_NatCom14}, while ARPES studies proposed that superconductivity sets in either the $\pi^*$ bands or the IL band alone~\cite{Sugawara_NatPhys09,Valla_PRL09}. This debate was re-ignited by the experimental observation of a free-electron IL band in Ca- and Rb-intercalated bilayer graphene~\cite{Kanetani_PNAS12,Kleeman_PRB13,Kleeman_JPSJ14}. The possibility of superconductivity in C$_6$CaC$_6$ was suggested theoretically based on the analogy with bulk CaC$_6$ \cite{Mazin_PML10, Jishi_ASTP11}, however predictive {\it ab initio} calculations of the critical temperature have not yet been reported.

In this work we elucidate the role of the electron-phonon interaction in the normal and in the superconducting state of C$_6$CaC$_6$ by performing state-of-the-art {\it ab initio} calculations powered by electron-phonon Wannier interpolation~\cite{giustino_wannier,margine_eliashberg}. For the normal state we study the electron self-energy and spectral function in the Migdal approximation; for the superconducting state we solve the 
anisotropic Migdal-Eliashberg equations including Coulomb interactions from first principles. Our main findings are: (i)~superconductivity in C$_6$CaC$_6$ can be explained by a phonon-mediated pairing mechanism; (ii)~in contrast to bulk CaC$_6$, low-energy Ca vibrations are responsible for the majority of the EPC in the superconducting state; (iii)~unlike bulk CaC$_6$, ~C$_6$CaC$_6$ should exhibit two superconducting gaps. For clarity all technical details are described in the Methods.

\begin{figure*}[t]
\begin{center}
\includegraphics[width=0.95\textwidth]{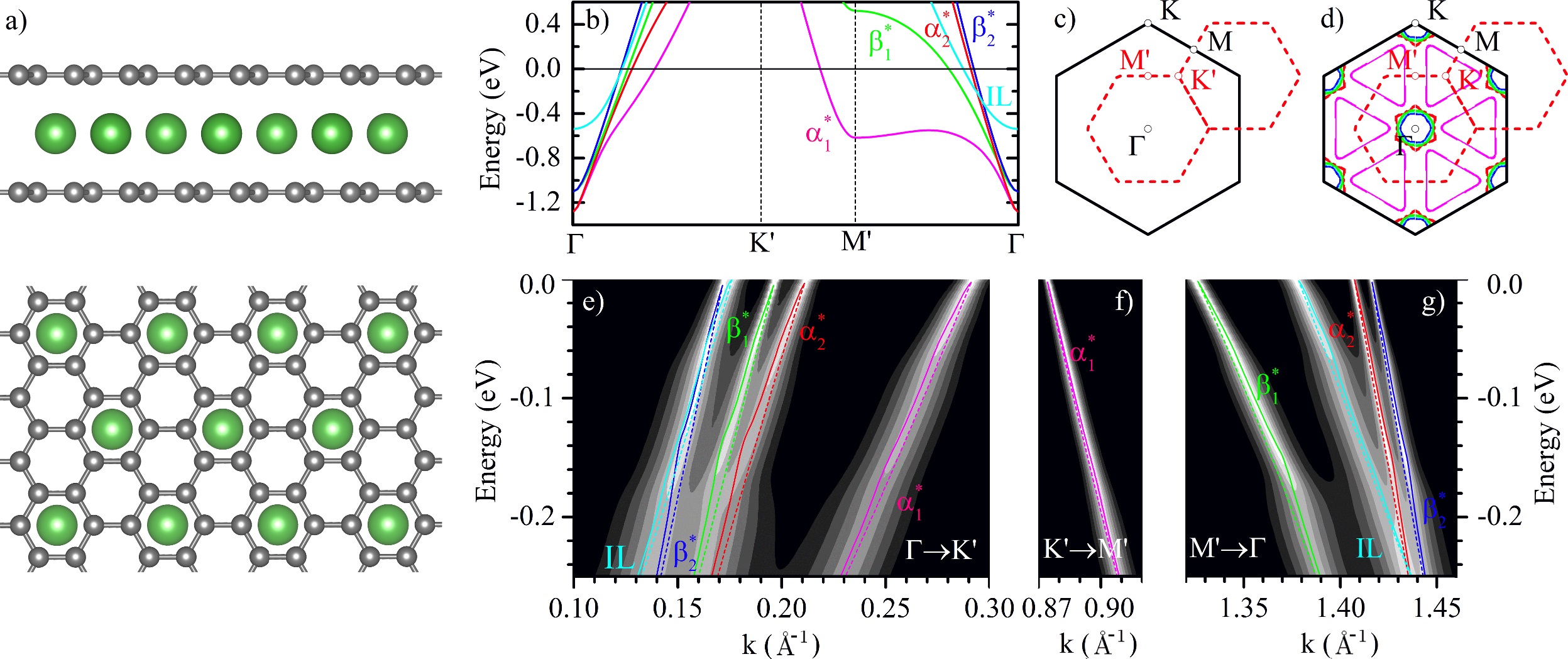}
\end{center}
\caption{\textbf{Crystal structure, band dispersion, Fermi surface, and spectral function of bilayer C$_6$CaC$_6$.}
  (a) Side- and top-view of a ball-and-stick model of C$_6$CaC$_6$, with C in grey and Ca in green. The structure is analogous to bulk CaC$_6$ \cite{Emery_PRL05}. (b) Band structure of C$_6$CaC$_6$. The outermost   $\pi^*$ bands (with respect to $\Gamma$ point) are labeled as $\alpha_1^*$ (magenta line) and $\beta_1^*$   (green line), the innermost $\pi^*$ bands are labeled as $\alpha_2^*$ (red line) and $\beta_2^*$ (blue lines).   The interlayer band is labeled as IL (cyan line). (c) Brillouin zones of a graphene unit cell (black   full lines) and a $\sqrt{3}\times\sqrt{3} \,R30^{\circ}$ graphene supercell (red dashed lines). (d) Two-dimensional Fermi surface sheets of C$_6$CaC$_6$, with the same color code as in (b). (e)-(g) Calculated spectral function of bilayer C$_6$CaC$_6$ in the normal state, along the same high-symmetry directions as shown in (b). The band structures with (solid lines) or without EPC (dashed lines) are overlaid on top of the spectral function. 
 }
\label{fig1}
\end{figure*}


Figure~\ref{fig1}(a) shows a ball-and-stick model of Ca-intercalated bilayer graphene, while Figs.~\ref{fig1}(b)-(d) show the corresponding band structure, Brillouin zone, and Fermi surface, respectively. Two sets of bands cross the Fermi level around the $\Gamma$ point. The bands labeled as $\alpha^*$ and $\beta^*$ in Fig.~\ref{fig1}(b) represent $\pi^*$ states, and are obtained by folding the Dirac cone of graphene from K to $\Gamma$, following the superstructure induced by Ca. The band labeled as IL is the Ca-derived nearly-free electron band, which disperses upwards from about 0.5~eV below the Fermi energy at~$\Gamma$~\cite{Kanetani_PNAS12,Mazin_PML10,Jishi_ASTP11}. In Fig.~\ref{fig1}(d) we see two sets of Fermi surface sheets: triangular hole pockets around the K$^\prime$ points, corresponding to the $\alpha_1^*$ states; and a bundle of small electron pockets around $\Gamma$, which arise from the other $\pi^*$ states ($\alpha_2^*$, $\beta_1^*$, and $\beta_2^*$) and from the IL band.

In order to quantify the strength of EPCs for each of these bands we calculate the spectral function in the normal state using the Migdal approximation~\cite{giustino2008N}. Figs.~\ref{fig1}(e)-(g) show that the EPC induces sudden changes of slope in the bands, which are referred to as `kinks' in high-resolution ARPES experiments. A pronounced kink is seen in all bands at a binding energy of 180~meV. This feature can be assigned to the coupling with the in-plane C$_{xy}$ stretching modes \cite{Park2008} (see Supplementary Fig.~S1 for the phonon dispersion relations). The occurrence of this high-energy kink in the $\pi^*$ bands is consistent with the observed broadening of the quasiparticle peaks in the ARPES spectra of bulk CaC$_6$ in the same energy range~\cite{Yang_NatCom14}. This feature results from a sharp peak at 180~meV in the real part of the electron self-energy of the $\pi^*$ electron pockets, as shown in Supplementary Fig.~S2. A second kink is clearly seen at a binding energy of 70~meV, and is mostly visible for the $\pi^*$ bands defining the hole pockets [magenta line in Fig.~\ref{fig1}(e)] and for the interlayer band [cyan line in Fig.~\ref{fig1}(g)]. This low-energy kink corresponds to a second, smaller peak at the same energy in the real part of the electron self-energy (Supplementary Fig.~S2), and arises from the coupling with the out-of-plane C$_{z}$ modes of the graphene sheets. Closer inspection reveals also a third kink around 12~meV, however this feature is hardly discernible and is unlikely to be observed in ARPES experiments. This faint structure arises from the coupling to the in-plane Ca$_{xy}$ vibrations, and can be seen more clearly in the real and the imaginary parts of the electron self-energy in Supplementary Fig.~S2.

From the calculated electron self-energy we can extract the electron-phonon mass enhancement parameter $\lambda_{\rm F}$ for each band using the ratio  between the bare and the renormalized Fermi velocities. Along the $\Gamma$K$^\prime$ direction we obtain $\lambda_{\rm F}= 0.53$ ($\alpha_1^*$), 0.48 ($\alpha_2^*$), 0.30 ($\beta_1^*$ and $\beta_2^*$) for the $\pi^*$ bands, and $\lambda_{\rm F}=0.68$ for the IL band. Along the other two directions $\Gamma$M$^\prime$ and M$^\prime$K$^\prime$ the mass enhancement parameters are up to a factor of three smaller than the corresponding values along $\Gamma$K$^\prime$, suggesting a rather anisotropic EPC. Our findings are in agreement with ARPES studies on bulk KC$_6$ and CaC$_6$ superconductors~\cite{Gruneis_PRB09,Valla_PRL09} and also with recent work on Rb-intercalated bilayer graphene~\cite{Kleeman_JPSJ14}. 

\begin{figure*}[t]
\begin{center}
\includegraphics[width=0.9\textwidth]{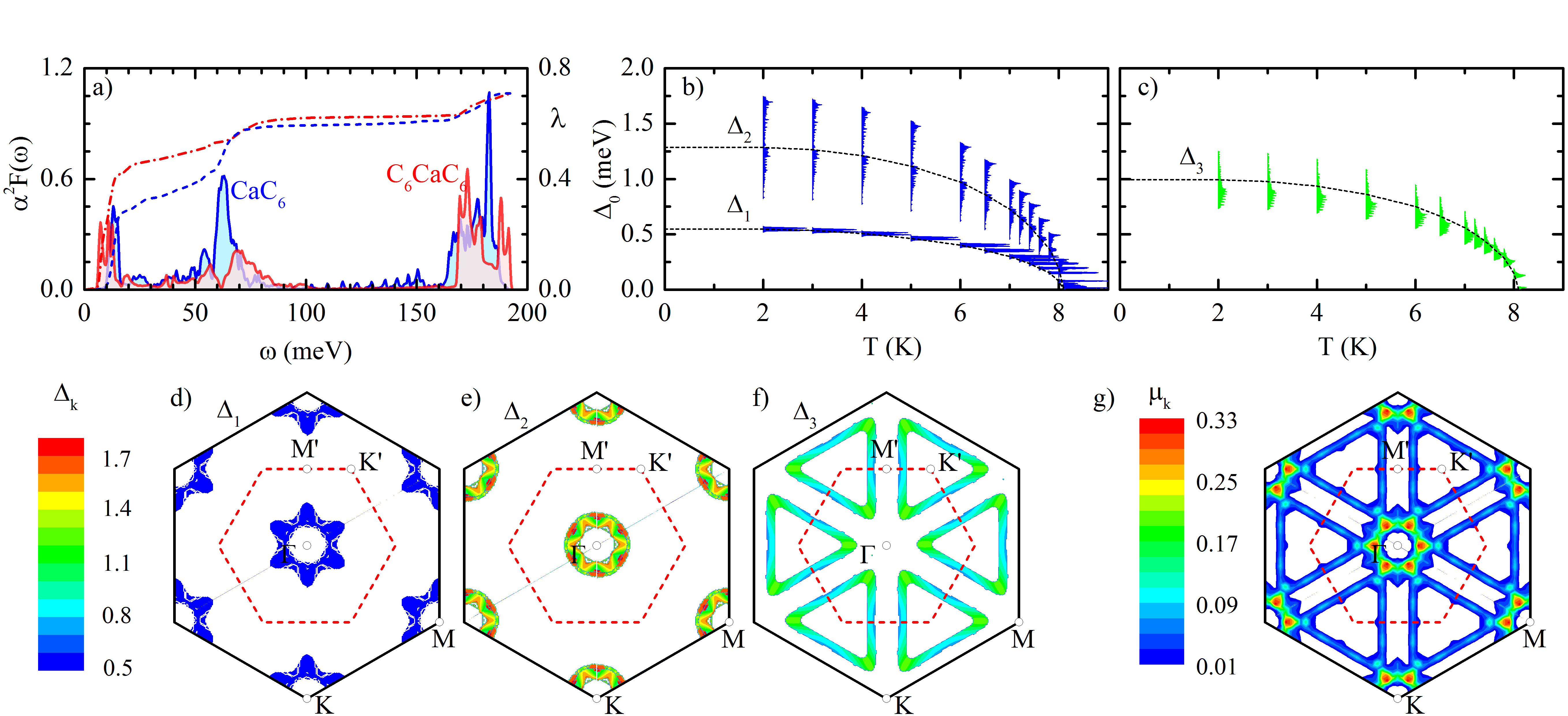}
\end{center}
\caption{\textbf{Electron-phonon coupling and superconducting gap function of bilayer C$_6$CaC$_6$.}
  (a) Eliashberg spectral function and cumulative EPC calculated for CaC$_6$ (blue) and C$_6$CaC$_6$ (red). The solid lines are for $\alpha^2F(\omega)$ (left scale), the dashed lines are for $\lambda(\omega)$ (right   scale). (b)-(c) Energy distribution of the anisotropic superconducting gaps $\Delta_\bbk$ of C$_6$CaC$_6$, centered around the $\Gamma$ and K$^\prime$ points as a function of temperature. The gaps were calculates using the {\it ab initio} Coulomb pseudopotential $\mu^*=0.155$. The dashed black lines indicate the average values of the gaps. The gaps vanish at the  critical temperature $T_{\rm c}=8.1$~K. The color-coded gaps at the lowest temperature refer to the  segments $\Delta_1$, $\Delta_2$, and $\Delta_3$ discussed in the text, and can approximately be identified with the panels (d), (e), and (f), respectively. (d)-(f) Momentum-resolved superconducting gap $\Delta_\bbk$ (in meV) on the Fermi surface at zero temperature: (d) and (e) correspond to the lower gap $\Delta_1$ and the upper gap $\Delta_2$ centered around the $\Gamma$ point,  (f) corresponds to the $\Delta_3$ centered around the K$^\prime$ point. (g) Dimensionless anisotropic Coulomb pseudopotential $\mu_\bbk$ on the Fermi  surface. For clarity in (d)-(g) the values correspond to electrons within $\pm 250$~meV from the Fermi energy (hence the `thick' Fermi surface sheets).
  }
\label{fig2}
\end{figure*}

We now move from the normal state to the superconducting state, and consider an electron-phonon pairing mechanism in analogy with bulk CaC$_6$. Figure~\ref{fig2}(a) shows a comparison between the isotropic Eliashberg spectral functions, $\afo$, and the cumulative total EPC, $\lambda(\omega)$, calculated for C$_6$CaC$_6$ and for bulk CaC$_6$. In both cases we can distinguish three main contributions to the EPC, to be associated with the Ca$_{xy}$ vibrations ($\sim$10~meV), the out-of-plane C$_z$ vibrations ($\sim$70~meV), and the in-plane C$_{xy}$ modes ($\sim$180~meV). The Eliashberg functions of C$_6$CaC$_6$ and CaC$_6$ look similar in shape, and the total EPC is $\lambda=0.71$ in both cases. However the relative contributions of each set of modes differ considerably. The low-energy Ca modes are slightly softer in C$_6$CaC$_6$, leading to a larger contribution to the EPC than in CaC$_6$. These modes account for 60\% of the total coupling in the bilayer, while they only account for less than 30\% of the coupling in bulk CaC$_6$. In both cases the EPC strength associated with the in-plane C-C stretching modes is too weak (15\% of the total) to make a sizable contribution to the superconducting pairing. This is somewhat counterintuitive, given that the C$_{xy}$ modes lead to the most pronounced kinks in the spectral function in Fig.~\ref{fig1}. 

The smaller contribution of the out-of plane C$_z$ modes to the EPC and the softening of the in-plane Ca$_{xy}$ vibrations, obtained when going from the bulk to the bilayer, are similar to the results found for Li- and Ca-decorated monolayer graphene~\cite{Profeta_NatPhys12}. In the monolayer case, the removal of quantum confinement causes a shift of the IL wave function farther away from the graphene layer as compared to bulk, and, therefore, the EPC coupling between $\pi^*$ and IL states mediated by C$_z$ vibrations is reduced. Although in the bilayer case the IL state is strongly localized around the Ca atom, the fact that the interlayer charge density is only present between the graphene layers gives rise to a weaker coupling of the  out-of-plane C$_z$ modes with the interlayer electrons, and, therefore, a lower contribution to the global EPC. 

To check whether the softening of the low energy Ca$_{xy}$ modes is due to the difference in the structural parameters between the bilayer and the bulk, we recalculate the vibrational spectrum of bilayer C$_6$CaC$_6$ after setting the in-plane lattice constant and the interlayer distance to the values of the bulk. As shown in Supplementary Fig.~S3, the in-plane C$_{xy}$ modes ($\omega > 105$~meV) soften due to the increase in the in-plane lattice constant, and the out-of-plane C$_z$ and Ca$_z$ modes ($20 < \omega < 45$~meV ) harden  due to the decrease in the interlayer distance. On the other hand, the two lowest-lying modes involving mainly in-plane Ca$_{xy}$ vibrations are considerably less affected, although a closer inspection reveals  an overall slight hardening of approximately 1~meV, particularly along the $\Gamma$M$^\prime$ and M$^\prime$K$^\prime$ directions. This suggests that the observed softening of the low-energy Ca$_{xy}$ modes from bulk to bilayer is most likely caused by changes in the electronic structure which is consistent with an increased density of states at the Fermi level. 

In order to determine the superconducting critical temperature of C$_6$CaC$_6$ we solve the anisotropic Eliashberg  equations~\cite{allen_mitrovic,margine_eliashberg,margine_graphene}, with the Coulomb pseudopotential $\mu^*$ calculated from first principles (the calculation of $\mu^*$ is discussed below). In Figs.~\ref{fig2}(b)-(c) we plot the energy-dependent distribution of the superconducting gap, separated into contributions corresponding to the two sets of Fermi surfaces centered around the $\Gamma$ and K$^\prime$ points. We see that two distinct gaps open on the $\Gamma$-centered electron pockets, with average values $\Delta_1=0.55$~meV and $\Delta_2=1.29$~meV at zero temperature.  The $\Delta_1$ gap is characterized by a very narrow energy profile and the EPC on these pockets is essentially isotropic, resulting primarily from the coupling with the Ca$_{xy}$ phonons [Fig.~\ref{fig2}(d)]. The $\Delta_2$ gap exhibits a much broader energy profile ($0.82\!<\!\Delta_{2}\!<\!1.75$~meV) and originates mainly from the coupling to Ca modes and out-of-plane C$_{z}$ phonons [Fig.~\ref{fig2}(e)].  In between the two $\Gamma$-centered gaps, a third gap with an average value $\Delta_3=0.99$~meV opens on the triangular hole pockets around K$^\prime$ [$\alpha_1^*$ states] [Fig.~\ref{fig2}(f)]. These states couple primarily to C$_{xy}$ phonons and the gap has an anisotropic character with a large spread in energy ($0.73<\Delta_{3}<1.25$~meV). In Supplementary Fig.~S4 we show the anisotropic EPC parameters $\lambda_\bbk$ leading to this superconducting gap structure. For completeness, we compare our results with the gap structure of bulk CaC$_6$. In the latter case, only one superconducting gap is predicted (see Supplementary Fig.~S5), in agreement with previous theoretical studies~\cite{Sanna_PRB07,Sanna_PRB12} and experiments~\cite{Bergeal_PRL06,Lamura_PRL06}. Although multiple sheets of the Fermi surface contribute to the superconducting gap, there is no separation into distinct gaps, giving rise to a smeared multigap structure~\cite{Sanna_PRB07,Sanna_PRB12}. This situation is similar to the $\Delta_2$ and $\Delta_3$ gaps in the bilayer case. Based on our results we suggest that in C$_6$CaC$_6$ high-resolution ARPES experiments might be able to resolve two distinct gaps on the electron pockets, corresponding to $\Delta_1$ and $\Delta_{2}$, but only one gap on the triangular hole pockets, corresponding to $\Delta_{3}$.

The calculations of the gap function and the superconducting critical temperature in Fig.~\ref{fig2}(b)-(c) require the knowledge of the Coulomb pseudopotential $\mu^*$. In order to determine this parameter we first calculate the dimensionless electron-electron interaction strength within the random-phase approximation, obtaining $\mu=0.254$ (see Methods). Then we renormalize this interaction using $\mu^* = \mu/[1+\mu\log(\omega_{\rm pl}/\omega_{\rm ph})]$ \cite{morel62}, where $\omega_{\rm pl}$ and $\omega_{\rm ph}$ are characteristic electron and phonon energy scales, respectively. We set $\omega_{\rm pl}=2.5$~eV, corresponding to the lowest plasmon resonance in GICs \cite{mele79,echeverry12}, and $\omega_{\rm ph}=200$~meV, corresponding to the highest phonon energy in C$_6$CaC$_6$. Figure~\ref{fig2}(g) shows the variation of the Coulomb pseudopotential $\mu_{\bbk}$ across the Fermi surface. The repulsive interaction is strongest on the electron pockets, and weakest on the hole pockets. Since the EPC exhibits a very similar anisotropy, the net coupling strength $\lambda_\bbk - \mu^*_\bbk$ is only moderately anisotropic and can be replaced by a single, average $\mu^*$ in the Eliashberg equations. From Fig.~\ref{fig2}(g) we obtain $\mu^*=0.155$, and this is the value employed in Figs.~\ref{fig2}(b)-(c). As we can see in Figs.~\ref{fig2}(b)-(c), our Eliashberg calculations yield a superconducting critical temperature $T_{\rm c}=8.1$~K, which is only slightly higher than the experimental value $T_{\rm c}^{\rm exp}=4$~K reported in Ref.~\citenum{Hasegawa}. 

In order to check the role of anisotropic Coulomb interactions we repeat the calculations by considering a Coulomb pseudopotential resolved over the electron and hole pockets. In this case we find the decrease in $T_{\rm c}$ to be very small, $\Delta T_{\rm c}=0.3$~K. Furthermore we explore the sensitivity of the calculated $T_{\rm c}$ to the choice of the characteristic phonon energy $\omega_{\rm ph}$. To this aim we solve the Eliashberg equations again, this time by setting $\omega_{\rm ph}$ equal to the Matsubara frequency cutoff ($5\times200$~meV, see Methods). This alternative choice leads to $\mu^*=0.207$ and $T_{\rm c}=6.8$~K (Supplementary Fig.~S6), which is in even better agreement with experiment. For completeness in Supplementary Fig.~S7 we also show the dependence of $T_{\rm c}$ on the characteristic energy $\omega_{\rm c}$ as obtained using the standard McMillan formula~\cite{McMillan}. Consistent with our Eliashberg calculations, we find that large variations of $\omega_{\rm c}$ only change $T_{\rm c}$ by a few K's. These additional tests show that our results are solid, therefore we can safely claim that the {\it ab initio} Eliashberg theory yields $T_{\rm c}=6.8$--8.1~K for C$_6$CaC$_6$. The close agreement between these values and experiment supports the notion that Ca-intercalated bilayer graphene is a conventional phonon-mediated superconductor. 

In conclusion, we studied entirely from first principles the electron-phonon interaction and the possibility of phonon-mediated pairing in the newly-discovered superconducting C$_6$CaC$_6$. We showed that the Ca vibrations play an important role in the pairing but do not carry a sharp signature in the normal-state band structure; conversely the high-frequency in-plane C vibrations lead to pronounced photoemission kinks but have a small contribution to the pairing. The good agreement between the critical temperature calculated here and the recent experiments of Ref.~\citenum{Hasegawa} indicate that Ca-intercalated bilayer graphene is an electron-phonon superconductor. The present work calls for high-resolution spectroscopic investigations, as well as for calculations based on alternative {\it ab initio} methods \cite{scdft1,scdft2}, in order to test our prediction of two distinct superconducting gaps in C$_6$CaC$_6$.

\section*{Methods}
 
The calculations are performed within the local density approximation to density-functional theory~\cite{lda1,lda2} using planewaves and norm-conserving pseudopotentials~\cite{nc1,nc2}, as implemented in the {\tt Quantum-ESPRESSO} suite~\cite{QE}. The planewaves kinetic energy cutoff is 60~Ry and the structural optimization is performed using a threshold of 10~meV/\AA\ for the forces. C$_6$CaC$_6$ is described using the $\sqrt{3}\times\sqrt{3}$~$R30^{\circ}$ supercell 
of graphene with one Ca atom per unit cell, and periodic images are separated by 15~\AA. The optimized in-plane lattice constant and interlayer separation are $a=4.24$~\AA\ and $d=4.50$~\AA, respectively. Bulk CaC$_6$ is described using the rhombohedral lattice, and the optimized lattice constant and rhombohedral angle are $a=5.04$~\AA\ and $\alpha=50.23^{\circ}$, respectively. The electronic charge density is calculated using an unshifted Brillouin zone mesh with $24^2$ and $8^3$ $\bbk$-points for C$_6$CaC$_6$ and CaC$_6$, respectively, and a Methfessel-Paxton smearing of 0.02~Ry. The dynamical matrices and the linear variation of the self-consistent potential are calculated within density-functional perturbation theory~\cite{baroni2001} on the irreducible set of regular $6^2$ (C$_6$CaC$_6$) and $4^3$ (CaC$_6$) $\bbq$-point grids. 

The electron self-energy, spectral function, and superconducting gap are evaluated using the {\tt EPW} code~\cite{giustino_wannier,EPW,margine_eliashberg}. The electronic wavefunctions required for the Wannier-Fourier interpolation~\cite{marzari_rmp,wannier} in {\tt EPW} are calculated on uniform unshifted Brillouin-zone grids of size $12^2$ (C$_6$CaC$_6$) and $8^3$ (CaC$_6$). The normal-state self-energy is calculated on fine meshes consisting of 100,000 inequivalent $\bbq$-points, using a broadening parameter of 10~meV and a temperature $T=20$~K. For the anisotropic Eliashberg equations we use 120$\times$120 (40$\times$40$\times$40) $\bbk$-point grids and 
60$\times$60 (20$\times$20$\times$20) $\bbq$-point grids for C$_6$CaC$_6$ (CaC$_6$). The Matsubara frequency cutoff is set to five times the largest phonon frequency ($5\times 200$~meV), and the Dirac delta functions are replaced by Lorentzians of widths 100~meV and 0.5~meV for electrons and phonons, respectively. The technical details of the Eliashberg calculations are described extensively in Ref.~\citenum{margine_eliashberg}.
 
The electron-electron interaction strength is obtained as $\mu=N_{\rm F}\langle\langle V_{\bbk,\bbk'} 
\rangle\rangle_{\rm FS}$, where $N_{\rm F}$ is the density of states at the Fermi energy,
$V_{\bbk,\bbk'}=\langle \bbk,-\bbk | W | \bbk',-\bbk'\rangle$, and $W$ is the screened Coulomb interaction within the random phase approximation \cite{mucohen}. Here $\langle\langle \cdot \rangle\rangle_{\rm FS}$ indicates a double Fermi-surface average, and $\bbk$ stands for both momentum and band index. The screened Coulomb interaction is calculated using the Sternheimer approach~\cite{sternheimer,giustino2010}. Linear-response equations are solved using 26$\times$26 Brillouin-zone grids, corresponding to 70~inequivalent points. The energy cutoff of the dielectric matrix is 10~Ry (815 planewaves). 

H.L. and F.G. acknowledge support from the European Research Council (EU FP7/grant no. 604391 Graphene Flagship and EU FP7/ERC grant no. 239578).


\end{document}


\title{Electron-phonon interaction and pairing mechanism in superconducting Ca-intercalated bilayer graphene}
    
\author{E. R. Margine}
\affiliation{Department of Physics, Applied Physics and Astronomy, 
Binghamton University - SUNY, Binghamton, New York 13902, USA}

\author{Henry Lambert}
\affiliation{Department of Materials, University of Oxford, Parks Road,
Oxford OX1 3PH, United Kingdom}

\author{Feliciano Giustino} 
\affiliation{Department of Materials, University of Oxford, Parks Road,
Oxford OX1 3PH, United Kingdom}

\maketitle

\begin{figure}[h]
\vspace{1cm}
\begin{center}
\includegraphics[width=0.5\textwidth]{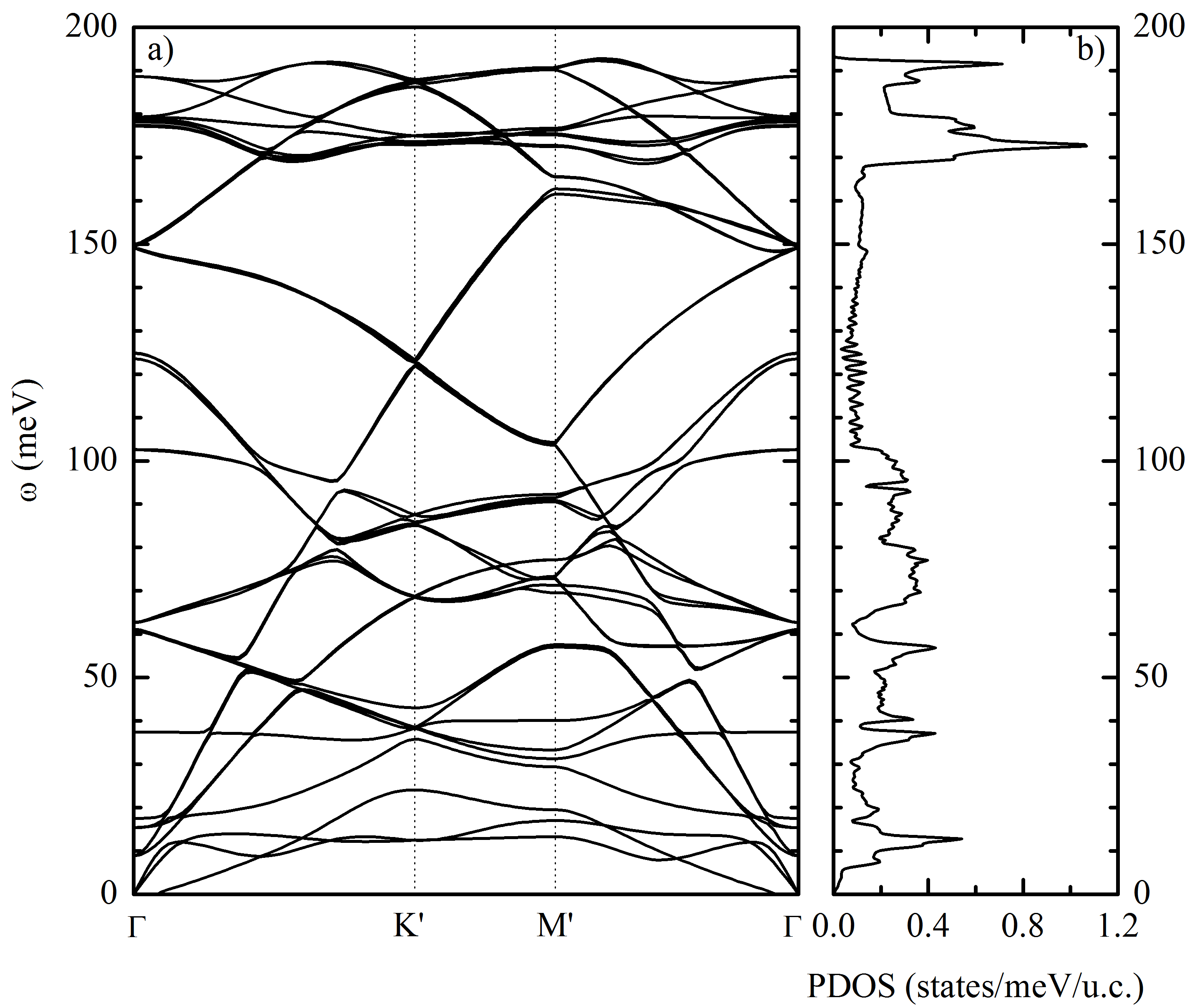}
\end{center}
\caption{\textbf{Calculated phonon dispersion and phonon density of states for C$_6$CaC$_6$.}
 (a) phonon dispersion relations and (b) phonon density of states.}
\end{figure}

\begin{figure}[h]
\begin{center}
\includegraphics[width=0.6\textwidth]{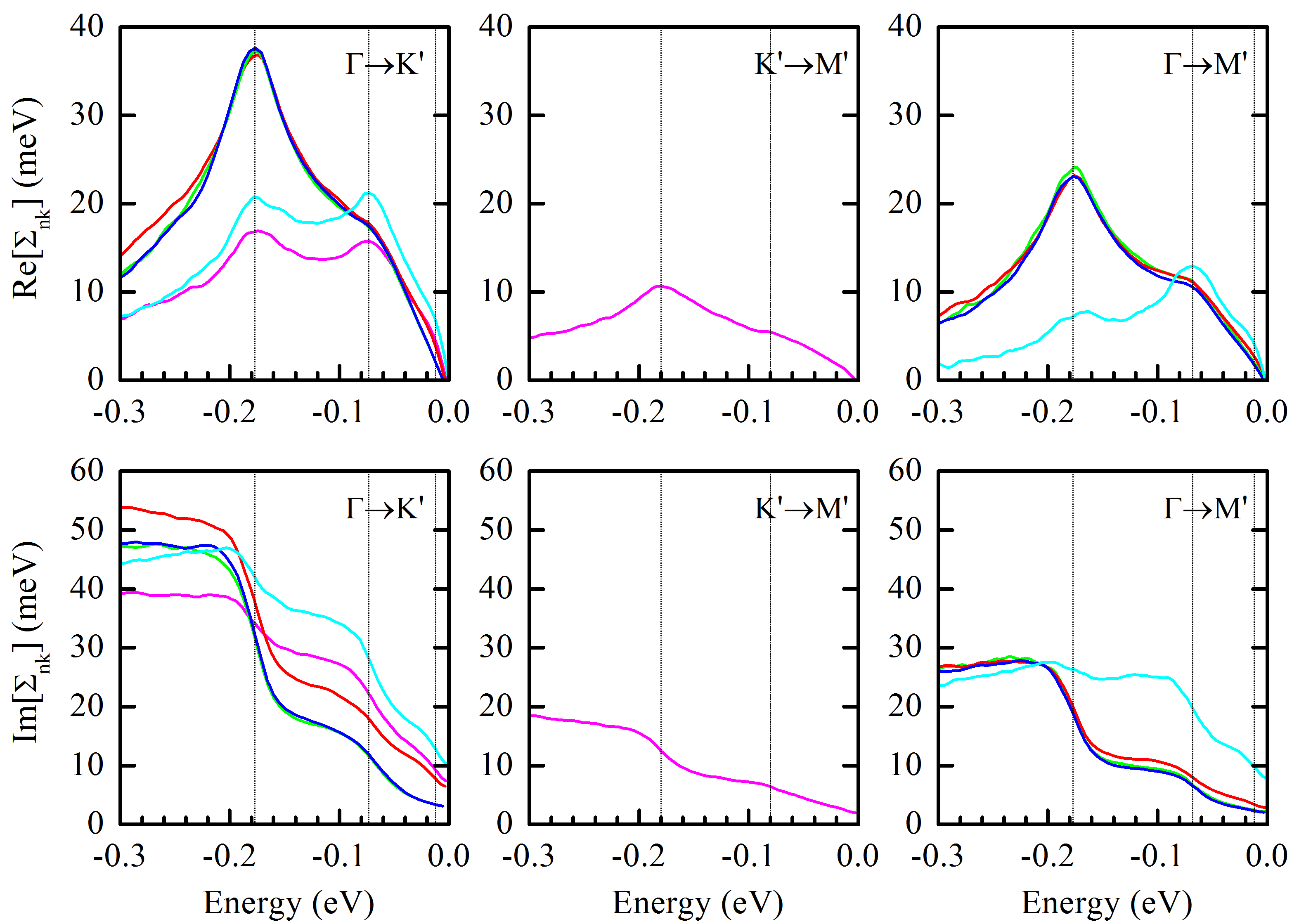}
\end{center}
\caption{\textbf{Calculated electron-self energy for C$_6$CaC$_6$.}
Real part (top row) and imaginary part (bottom row) of the electron self-energy $\Sigma$ arising from the 
electron-phonon interaction, calculated for Ca-intercalated bilayer graphene in the normal state.
The color code is the same as in Fig.~1(b) of the main text:
for the $\pi^*$ bands magenta is for $\alpha_1^*$, green for $\beta_1^*$, red for $\alpha_2^*$, and blue
for $\beta_2^*$. The interlayer band is in cyan. The high-symmetry directions in the Brillouin zone
are indicated in each panel and correspond to Fig.~1(c) of the main text. The vertical grey lines
indicate the vibrational frequencies of the three prominent features at 12~meV, 70~meV, and 180~meV
discussed in the main text.
}
\end{figure}

\begin{figure}[h]
\vspace{1cm}
\begin{center}
\includegraphics[width=0.5\textwidth]{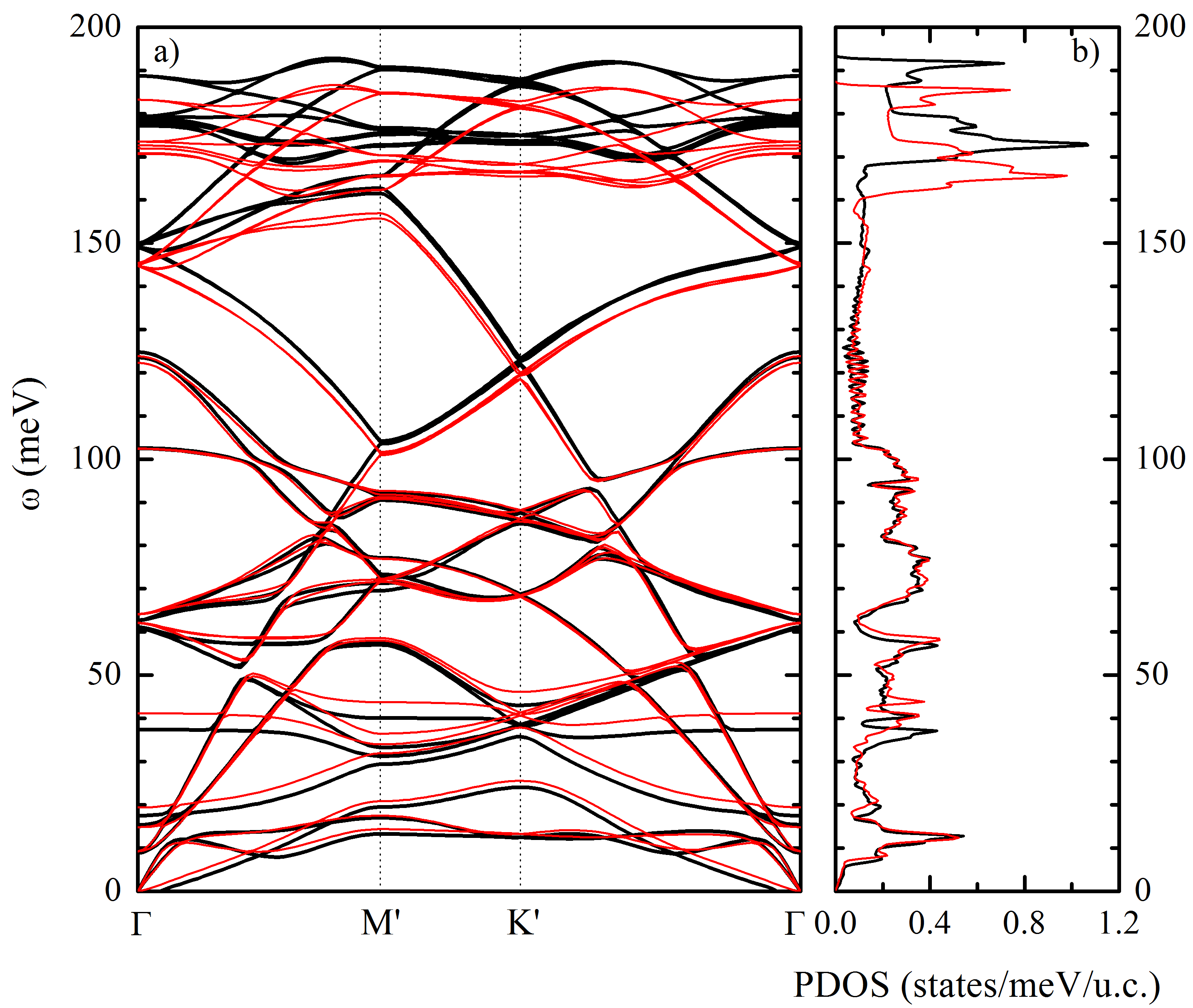}
\end{center}
\caption{\textbf{Calculated phonon dispersion and phonon density of states for C$_6$CaC$_6$.}
(a) phonon dispersion relations and (b) phonon density of states.
Black lines correspond to the relaxed structural parameters of bilayer C$_6$CaC$_6$ ($a=4.2429$ \AA\, and $c=4.5027$ \AA\,) and red lines correspond to the relaxed structural parameters of bulk CaC$_6$ ($a=4.2784$ \AA\, and $c=4.3937$ \AA\,).}
\end{figure}

\begin{figure}[h]
\begin{center}
\includegraphics[width=0.8\textwidth]{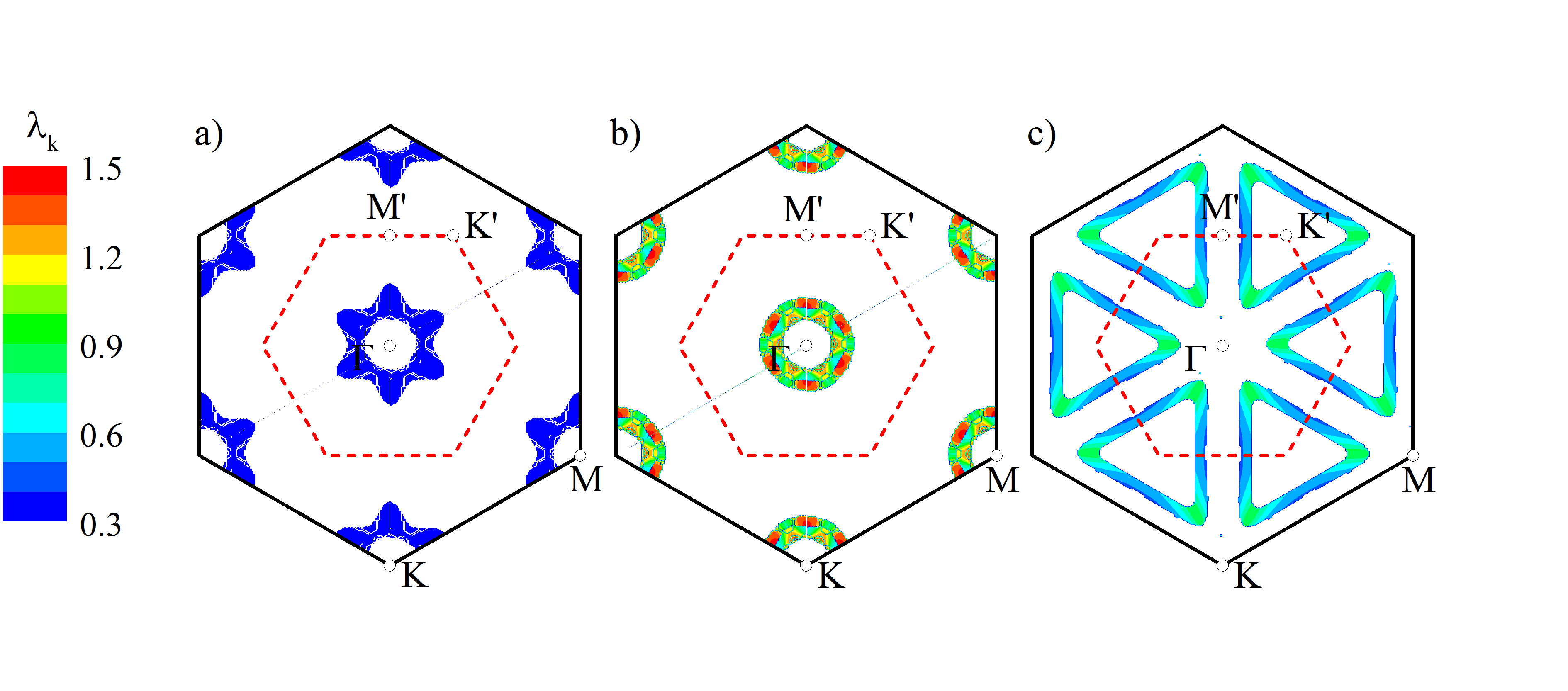}
\end{center}
\caption{
\textbf{Calculated momentum-resolved electron-phonon coupling parameters $\lambda_{\bf k}$ on the Fermi surface of Ca-intercalated bilayer graphene.} These are the EPC values employed in the solution of the anisotropic Eliashberg equations leading to Figs.~2(d)-(f) in the main text. As in the main text, the data points correspond to electrons within $\pm 250$~meV from the Fermi energy (this explains why we have `thick' Fermi surface sheets as opposed to thin lines).
}
\end{figure}

\begin{figure}[h]
\begin{center}
\includegraphics[width=0.55\textwidth]{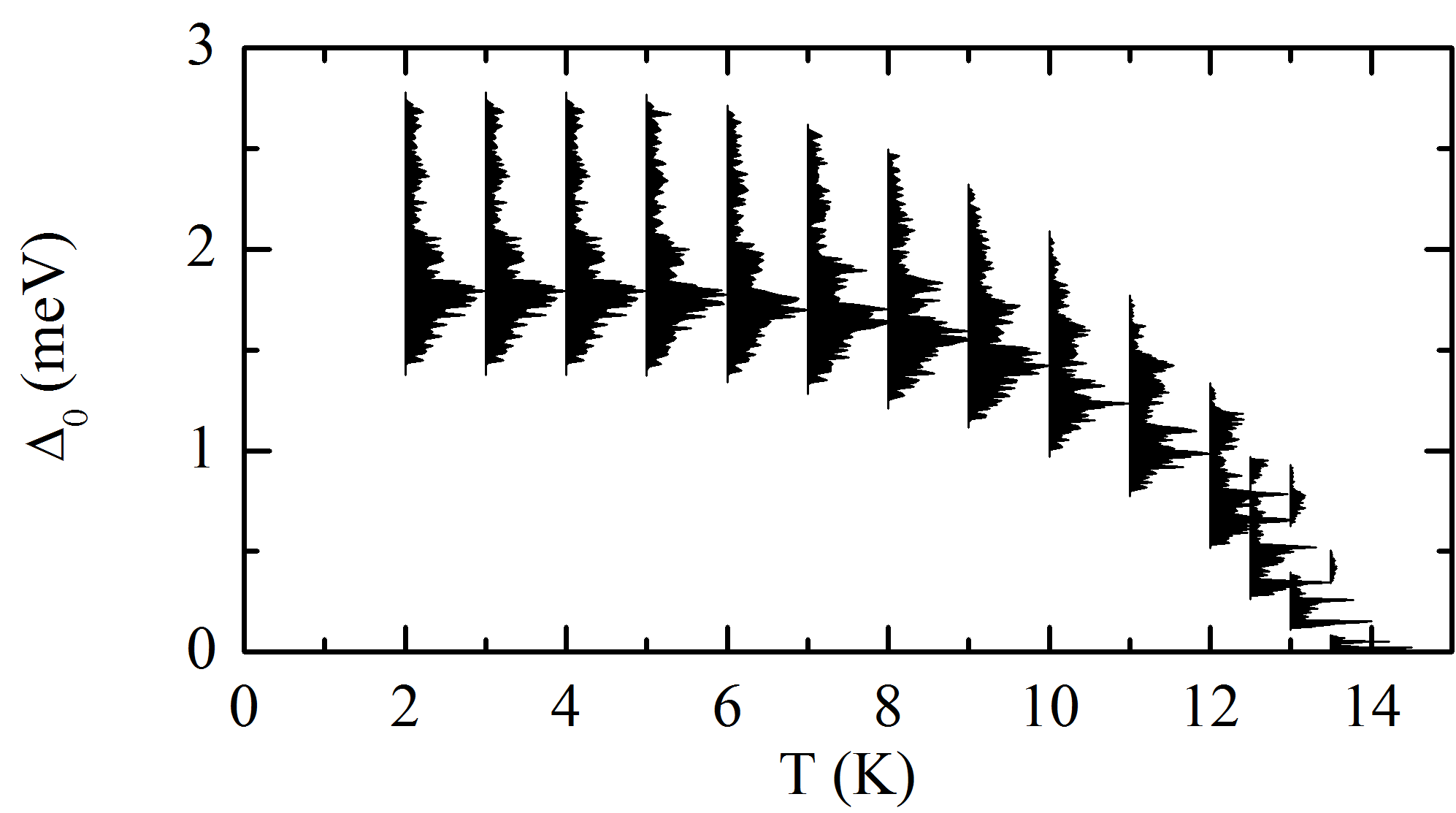}
\end{center}
\caption{
\textbf{Distribution of the superconducting gap function calculated for bulk CaC$_6$.} This is to be compared
with Figs.~2(b)-(c) of the main text. Unlike C$_6$CaC$_6$, in this case we do not find two distinct superconducting 
gaps. In this case the Eliashberg equations were solved using $\mu^*=0.14$, as calculated within the
random-phase approximation (see Methods in the main text). The calculated critical temperature 
$T_{\rm c}=13.5$~K is in good agreement with the measured value $T_{\rm c}^{\rm exp}=11.5$~K (Ref.~[11] of the main text).
}
\end{figure}

\begin{figure}[h]
\begin{center}
\includegraphics[width=\textwidth]{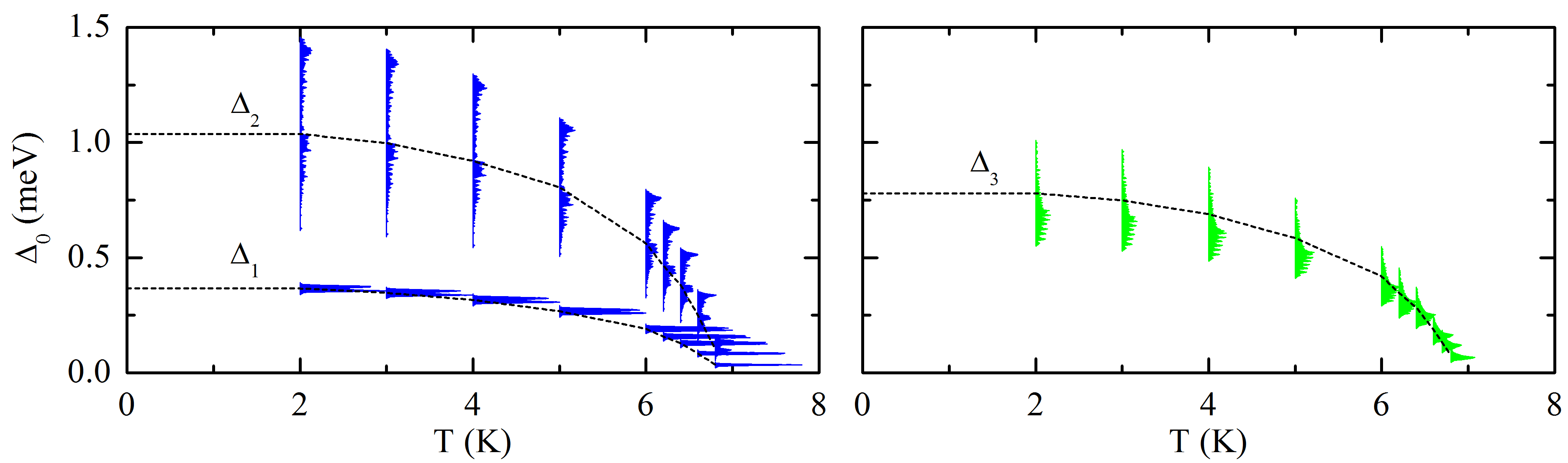}
\end{center}
\caption{
\textbf{Distribution of the superconducting gap function calculated for Ca-intercalated bilayer graphene.} The computational setup is the same as in Figs.~2(b)-(c) of 
the main text, except in this case we set the characteristic phonon energy $\omega_{\rm ph}$ to
the Matsubara frequency cutoff ($5\times200$~meV), leading to $\mu^*=0.207$.
In this case the calculated critical temperature is $T_{\rm c}=6.8$~K. The dashed lines are guides
to the eye. 
}
\end{figure}

\begin{figure}[h]
\begin{center}
\includegraphics[width=0.5\textwidth]{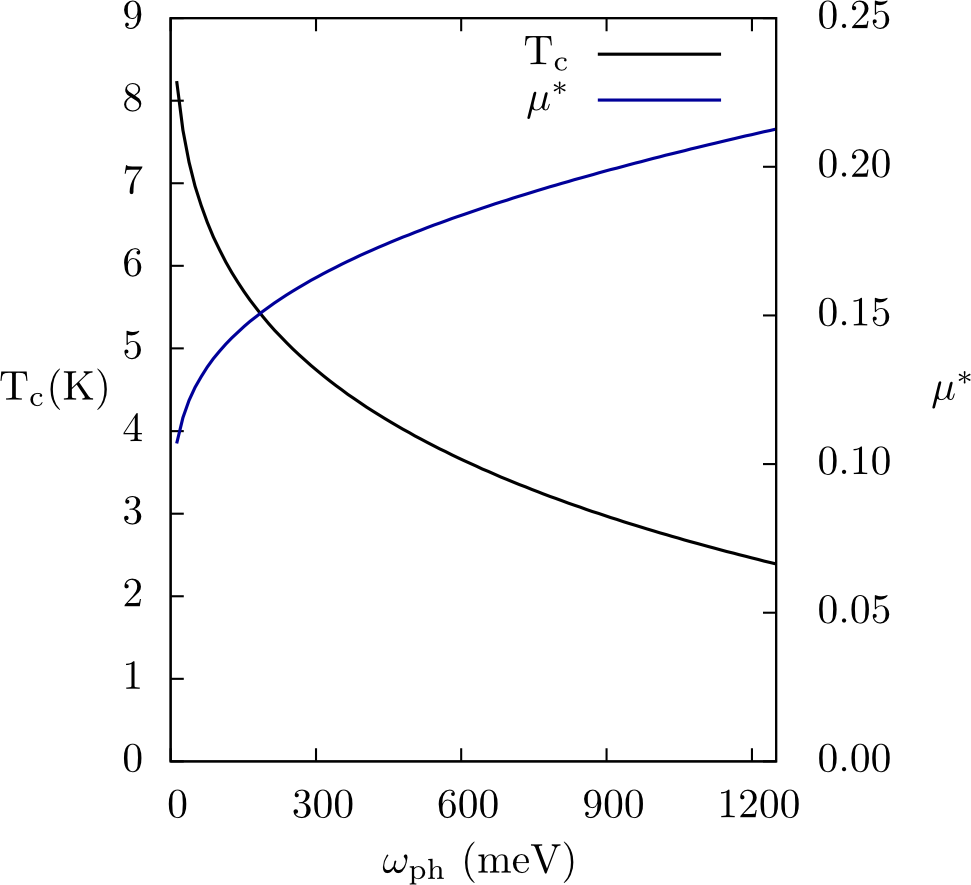}
\end{center}
\caption{\textbf{Dependence of $T_{\rm c}$ and $\mu^*$ on the characteristic phonon energy $\omega_{\rm c}$.}
Sensitivity of the calculated critical temperature of C$_6$CaC$_6$ to the choice of the
characteristic phonon energy $\omega_{\rm c}$ in $\mu^* = \mu/[1+\mu\log(\omega_{\rm pl}/\omega_{\rm ph})]$.
For simplicity we plot the critical temperature obtained by using the McMillan equation
$T_{\rm c} = \omega_{\rm log}/1.2 \exp\{-1.04(1+\lambda)/[\lambda-\mu^*(1+0.62 \lambda)]\}$, 
with $\omega_{\rm log}=21.1$~meV, $\lambda = 0.71$, and $\mu=0.254$. The variation of $T_{\rm c}$
between $\omega_{\rm c}=200$~meV (the highest phonon energy) and $\omega_{\rm c}=5\times 200$~meV
(the Matsubara frequency cutoff) is $\Delta T_{\rm c}=2.6$~K, in good agreement with our
{\it ab initio} Eliashberg calculations described in the main text (for which $\Delta T_{\rm c}=1.6$~K).
}
\end{figure}